\documentclass[a4paper,11pt]{article} 
\usepackage[margin=1in]{geometry} 
\usepackage[title]{appendix}
\usepackage{amssymb, mathrsfs}
\usepackage{amsmath,epsfig, epsf, epic, graphicx, fontenc} 
\usepackage{fancybox, lscape, subfigure} 
\usepackage{bm}
\usepackage{natbib}
\usepackage{physics}
\usepackage[margin=2cm]{caption}
\usepackage{setspace}
\usepackage{hyperref}
\usepackage[ruled,vlined]{algorithm2e}
\SetKwInput{KwInput}{Input}                
\SetKwInput{KwOutput}{Output}              
\SetKwProg{Initialize}{Initialize:}{}{}

\newcommand{\R}{\mathbb{R}}
\DeclareMathOperator{\KL}{D_{\text{KL}}}
\renewcommand{\parallel}{\,\|\,}
\newcommand{\given}{\,|\,}
\DeclareMathOperator*{\argmin}{\arg\,\min}
\DeclareMathOperator*{\E}{E}
\DeclareMathOperator{\elbo}{\mathscr{L}}
\newcommand{\altgiven}{;\,}

\newcommand{\bb}[1]{\boldsymbol{#1}}

\allowdisplaybreaks

\title{Robust Coordinate Ascent Variational Inference with Markov chain Monte Carlo simulations}
\author{Neil Dey, Emmett Kendall}
\date{}

\begin{document}
\maketitle

\begin{abstract}
Variational Inference (VI) is a method that approximates a difficult-to-compute posterior density using better behaved  distributional families. VI is an alternative to the already well-studied Markov chain Monte Carlo (MCMC) method of approximating densities. With each algorithm, there are of course benefits and drawbacks; does there exist a combination of the two that mitigates the flaws of both? We propose a method to combine Coordinate Ascent Variational Inference (CAVI) with MCMC. This new methodology, termed Hybrid CAVI, seeks to improve the sensitivity to initialization and convergence problems of CAVI by proposing an initialization using method of moments estimates obtained from a short MCMC burn-in period.  Unlike CAVI, Hybrid CAVI proves to also be effective when the posterior is not from a conditionally conjugate exponential family.
\end{abstract}

\section{Introduction}
In Bayesian statistics, computing a posterior distribution is of primary interest; however, finding a closed form expression for the posterior proves to be a difficult task \citep{blei2017}. The main hurdle comes with computing the normalizing constant, since integration (especially in more than one dimension) can be computationally expensive. One workaround to this problem is to use MCMC sampling. Although MCMC will eventually converge to the true posterior distribution, the rate of convergence may be too slow for some practical uses \citep{casella2004}. 

In hopes of finding a more computationally efficient approximation method than MCMC, VI was developed. The key distinction is that VI uses optimization methods instead of sampling methods to learn the underlying posterior distribution \citep{blei2017}.  Specifically, CAVI is a form of VI that implements the coordinate ascent algorithm for its optimization routine. Although CAVI converges quickly, CAVI is highly sensitive to initialization in general; thus, CAVI is rarely applied outside of posteriors lying in the conditionally conjugate exponential family, where this flaw is mitigated by the existence of closed-form optimal updates \citep{blei2017}. 

We thus see that CAVI is generally limited in its application and MCMC cannot be efficiently used on large datasets. In this paper, we propose a new method that combines CAVI with MCMC, addressing the drawbacks inherent to both of these techniques. This new methodology is computationally efficient, applicable to any family of posteriors, and robust to initialization, allowing better scaling and wider application than either of the two existing methods.

The remainder of this paper is structured as follows. In Section \ref{sec:back}, we give further background for MCMC and CAVI and examine other approaches to combining them. In Section \ref{sec:hybrid}, we introduce the new method for combining MCMC and CAVI. In Section \ref{sec:simStudy}, we study the performance of our new method in comparison with existing posterior approximation techniques by conducting a simulation study\footnote{Code for the simulation study can be found at \url{https://github.com/neil-dey/robust-vi-mcmc}}. Specifically, Section \ref{subsec:exp} focuses on applying these methods to a posterior that is analytically known, whereas Section \ref{subsec:nonexp} applies the methods to an intractable posterior distribution.

\section{Background}
\label{sec:back}
The primary goal of both MCMC and VI is to estimate a posterior distribution of latent variables, $\vb*{z}$, given observed data, $\vb*{x}$. Rather than directly estimating the posterior $p(\vb*{z}\given\vb*{x})$, VI considers a family of simpler \textit{variational distributions}, $q(\vb*{z}\given\boldsymbol \lambda)$, that are shaped by a set of \textit{variational parameters}, $\boldsymbol \lambda$, and aims to find the variational parameter $\boldsymbol \lambda^*$ that allows $q(\vb*{z}\given\boldsymbol \lambda^*)$ to be ``close to" $p(\vb*{z}\given\vb*{x})$ \citep{zhang2019} .  VI measures this ``closeness" with the Kullback-Leibler (KL) divergence:
\begin{equation*}
    \KL(q\parallel p) = \int_\R \log(\frac{q(x)}{p(x)}) q(x) \dd{x}.
\end{equation*}

In this paper, we work under the popular framework of minimizing the KL divergence over a \textit{mean-field} variational family, which assumes that the latent variables are mutually independent. That is,
\begin{equation*}
    q(\vb*{z}; \vb*{\lambda}) = \prod_{i=1}^m q_i(z_i; \vb*{\lambda}_i).
\end{equation*}
Hence, we approximate the posterior $p(\vb*{z}\given \vb*{x})$ with $q^*(\vb*{z};\vb*{\lambda}^*)$ given by
\begin{equation*}
    q^*(\vb*{z};\vb*{\lambda}^*
) = \argmin_{q\in\mathscr{D}} \KL(q(\vb*{z}; \vb*{\lambda}) \parallel p(\vb*{z}\given\vb*{x})),
\end{equation*}
where $\mathscr{D}$ is the mean-field variational family.  However, calculating the KL divergence requires calculating the unknown marginal with respect to the data,  $p(\vb*{x})$. Thus, VI instead maximizes the tractable-to-compute Evidence Lower Bound (ELBO) \citep{blei2017}, denoted $\elbo$, given by
\begin{equation*}
    \elbo(q) =  \E_q[\log p(\vb*{z}, \vb*{x})] - \E_{q}[\log q(\vb*{z})] .
\end{equation*} 

\begin{algorithm}[!ht]
\DontPrintSemicolon
  
  \KwInput{A model $p(\vb*{x}, \vb*{z})$, a data set $\vb*{x}$}
  \KwOutput{A variational density $q(\vb*{z}) = \prod_{j=1}^m q_j(z_j)$}
  \Initialize{Variational factors $q_j(z_j)$}{}
   \While{ELBO has not converged}
   {
   		\For{$j \in \{1, \ldots, m\}$}{
   		    Set $q_j(z_j) \propto \exp(\E_{-j}[\log p(z_j \given \vb*{z}_{-j}, \vb*{x}))$
   		}
   		Compute $\elbo(q) = \E[\log p(\vb*{z}, \vb*{x})] + \E[\log q(\vb*{z})]$
   }
    \Return $q(\vb*{z})$
\caption{CAVI algorithm given by \cite{blei2017}}
\label{fig:caviAlg}
\end{algorithm}

A common method of maximizing the ELBO is CAVI, shown in Algorithm \ref{fig:caviAlg}. CAVI is the usual coordinate ascent algorithm, treating the variational factors of $q$ as the coordinates of the input to $\elbo$. To see this more explicitly, we can rewrite the ELBO as a function of the variational parameters: 
\begin{equation*}
    \elbo(\vb*{\lambda}_1, \ldots, \vb*{\lambda}_n) = \int_{\R^m} \log(\frac{p(\vb*{z}, \vb*{x})}{\prod_{i=1}^m q_i(z_i\altgiven\vb*{\lambda}_i)}) \cdot \prod_{i=1}^m q_i(z_i \altgiven \vb*{\lambda}_i) \dd{z}.
\end{equation*}
The behavior of coordinate ascent is well studied; a proof of convergence under weak assumptions is presented in \cite{luenberger1973}, and it can be shown that coordinate ascent converges linearly for (at least locally) strictly convex objective functions with compact sublevel sets \citep{abatzoglou1982}.  CAVI is also well-studied, with the coordinate ascent update for $q_i$ available in closed form for conditionally conjugate exponential families \citep{blei2017}; this closed form update is shown in the for-loop of Algorithm \ref{fig:caviAlg}. 

Some computational and inferential advantages of CAVI are shown in  \cite{braun2010}; however, the results rely on the convexity of the objective function, which is rarely the case for the ELBO. Thus, \cite{blei2017} points out that in the case of non-convex optimization, CAVI is highly sensitive to the initialization of the $q_j(z_j; \vb*{\lambda}_j)$ factors---a significant drawback to CAVI. Furthermore, the literature does not typically apply CAVI to non-conditionally conjugate exponential family posteriors. Thus, alternatives to CAVI have been suggested. For example, \cite{zhang2019}, \cite{hoffman2013}, and \cite{titsias194} all focus on stochastic gradient descent (SGD) approaches to the optimization rather than coordinate ascent. Regardless of the techniques used to optimize the ELBO, there has also been work in exploiting the structures of certain models to increase the speed of convergence, such as in \cite{zhang2019}. 

Many approaches to combining VI and MCMC exist in the literature.  For example, \cite{mimno2012sparse} implements Gibbs sampling to approximate the expectations that cannot be computed analytically but are necessary for the SGD approach to VI. Similarly, \cite{salimans2015markov} focuses on the SGD approach to VI, with the insight of adding auxiliary variables and integrating MCMC steps into the posterior approximations. Additionally, \cite{de2013variational} does impressive work in combining MCMC and VI by using VI to improve the proposals generated by the Metropolis-Hastings algorithm.
The work in \cite{ruiz2019contrastive} is notable in its approach to combine MCMC and VI by applying MCMC to the initial variational distribution, thus proposing a new variational distribution (possibly of a different family). Furthermore, they develop a new divergence criterion since exactly computing the KL divergence with respect to an MCMC posterior estimate is not feasible. 

\section{Hybrid CAVI}

\label{sec:hybrid}
As mentioned in the introduction, a point of contention for CAVI is its sensitivity to initial values, as it is possible for the algorithm to only find local extrema. Hence, we propose a new algorithm, termed Hybrid CAVI, that is less sensitive to initialization than CAVI. The idea of Hybrid CAVI is to use the burn-in process from MCMC to inform the initialization to a CAVI routine. This will improve the inferential robustness of CAVI while maintaining its computational speed. 

Hybrid CAVI begins by performing MCMC as usual. However, the burn-in period is very short---much less than what is needed to approach the stationary distribution. We then continue for another few MCMC iterations and use this as an approximation, $\widehat{\pi}$, to the true posterior distribution. We then calculate the moments of $\widehat{\pi}$ and initialize the variational parameters using the method of moments---so that the moments of the initial variational distribution match those of $\widehat{\pi}$. 

By performing these few initial MCMC steps, the inputs to the subsequent CAVI routine are often close enough to the truth that CAVI can finish quickly and accurately. Therefore, the power of our method is that it directly addresses  the main pitfall of CAVI (sensitivity to initialization), while still maintaining a fast runtime.

\section{Simulation Study}
\label{sec:simStudy}
\subsection{Application to Conditionally Conjugate Exponential Families}
\label{subsec:exp}
We now examine the efficacy the Hybrid CAVI approach to estimating a posterior distribution. In order to compare CAVI,  MCMC, and Hybrid CAVI, we use the following simple setup. We generate data
$$\vb*{x}_1, \ldots, \vb*{x}_{100}\overset{iid}{\sim} N_2\qty(\mqty[27 \\ 13], \mqty[38 & 0.8 \\ 0.8 & 4])$$
with a prior distribution for the mean $\vb*{\mu}$ given as $\vb*{\mu} \sim N_2(\vb*{0}, 50\vb*{I})$. The true posterior is then given by
\begin{equation}\label{eq:truepost}
    p(\mu\given\vb*{x}) \sim N_2\qty(\mqty[27.230 \\ 12.991], \mqty[0.377 & 0.00793 \\ 0.00793 & 0.400])
\end{equation}
For the CAVI and Hybrid CAVI algorithms, let the variational factors $q_i(\mu_i\altgiven \vb*{\lambda}_i)$ be normal:
\begin{equation*}
    \mu_i \given \vb*{\lambda}_i \sim N(\lambda_{i, 1}, \lambda_{i, 2}).
\end{equation*}

\subsection*{CAVI with Analytic Expected Value}
Given the above setup, it is straightforward to directly calculate the expectation needed in the CAVI algorithm shown in Algorithm \ref{fig:caviAlg}; this drastically improves computation time.  Table \ref{fig:results1} displays the computation times and convergence results for the CAVI algorithm using closed form expected values. We mentioned that a drawback of CAVI is sensitivity to initialization; however, with closed form updates available in this simulation, initializations for the mean and variance of the initial variational distribution proved not to be an issue in this particular example. The motivation for the three initial conditions, $\{\bb{\lambda^1, \lambda^2, \lambda^3}\}$, will become clearer in the section that follows. Needless to say, this demonstration in which CAVI converges close to the true posterior (as can be confirmed by comparison to the true posterior from Equation \ref{eq:truepost} or by the small KL divergence) in a short amount of time is a strong argument in favor of CAVI. It is clear that in this case, where closed-form updates for CAVI are available, there is no need for Hybrid CAVI.

\begin{table}[!htb]
\begin{center}
    \begin{tabular}{|c|c|c|c|}
    \hline
 Initialization & Time (s) & Estimated Posterior & KL Div.\\
    \hline
$\mqty[\vb*{\lambda^1}_1\\\vb*{\lambda^1}_2]= \mqty[(10 , 1) \\ (10 , 1)]$ & 0.696 & $N_2\qty(\mqty[27.233 \\ 12.991], \mqty[0.375 & 0 \\ 0 & 0.0398])$ $\vphantom{\mqty[0\\0\\0]}$ & 70.359\\
$\mqty[\vb*{\lambda^2}_1\\\vb*{\lambda^2}_2]= \mqty[(25 , 1) \\ (10 , 1)]$  & 0.692 &  $N_2\qty(\mqty[27.233 \\ 12.991], \mqty[0.375 & 0 \\ 0 & 0.0398])$ $\vphantom{\mqty[0\\0\\0]}$ & 70.359 \\
$\mqty[\vb*{\lambda^3}_1\\\vb*{\lambda^3}_2]= \mqty[(10 , 1) \\ (20 , 1)]$ & 0.670 &  $N_2\qty(\mqty[27.233 \\ 12.991], \mqty[0.375 & 0 \\ 0 & 0.0398])$ $\vphantom{\mqty[0\\0\\0]}$ & 70.359\\
    \hline
\end{tabular}
\caption{Results from using CAVI with a closed form expression for the coordinate updates written into the program.}
\label{fig:results1}
\end{center}
\end{table}

\subsection*{CAVI with Numerical Optimization of ELBO}
The previous subsection used a closed form expression for the CAVI update of each variational parameter. However, it is rare in research to be working with easy, closed form densities, let alone having conjugate priors such as in our example. Indeed, we only have closed-form expressions for CAVI updates for conditionally conjugate exponential families. Hence, it may often be the case that CAVI is run identically to block-coordinate ascent on the ELBO (i.e. the variational parameter update is obtained through numerical optimization). However, using numerical optimization methods rather than the optimal closed-form update at each step lends itself to sensitivity to initialization of variational parameter values and also increases the computational burden of the algorithm. We thus repeat the previous experiment so that the ELBO is calculated using numerical integration and coordinate updates are optimized using a bounded Newton-Conjugate Gradient algorithm.
\begin{table}[!htb]
\begin{center}
    \begin{tabular}{|c|c|c|c|c|}
    \hline
Initialization & Time (s) & Estimated Posterior & KL Div. \\
    \hline
$\mqty[\vb*{\lambda^1}_1\\\vb*{\lambda^1}_2]= \mqty[(10 , 1) \\ (10 , 1)]$  & 557.2 & $N_2\qty(\mqty[10.765 \\ 10.709], \mqty[0.230 & 0 \\ 0 & 0.980])$ $\vphantom{\mqty[0\\0\\0]}$ & 36456 \\
$\mqty[\vb*{\lambda^2}_1\\\vb*{\lambda^2}_2]= \mqty[(25 , 1) \\ (10 , 1)]$ & 572.2 & $N_2\qty(\mqty[25.524 \\  10.774], \mqty[0.568 & 0 \\ 0 & 0.984 ])$ $\vphantom{\mqty[0\\0\\0]}$ & 1014\\
$\mqty[\vb*{\lambda^3}_1\\\vb*{\lambda^3}_2]= \mqty[(10 , 1) \\ (20 , 1)]$ & 712.4 & $N_2\qty(\mqty[10.392 \\ 19.002], \mqty[0.897 & 0 \\ 0 & 0.326])$ $\vphantom{\mqty[0\\0\\0]}$  & $\infty$\\
    \hline
\end{tabular}
\caption{Results from using CAVI with a numerical optimization method to maximize the ELBO.}
\label{fig:results2}
\end{center}
\end{table}

From Table $\ref{fig:results2}$, we confirm our hypothesis that CAVI can be sensitive to initialization. We chose the three initial values above because we viewed them as reasonable guesses  for a practitioner to have:
\begin{itemize}
	\item The practitioner has no intuition for the covariance structure, and so initializes the covariance to be the identity matrix
	\item The practitioner uses several different forms for the mean term: one in which the components are equal ($\vb*{\lambda^1}$), and two in which they are unequal ($\vb*{\lambda^2}$ and $\vb*{\lambda^3}$) 
\end{itemize}
Evidently, the different initializations display wildly different behaviors. We see that $\bb{\lambda^1}$ and $\bb{\lambda^3}$ provide convergence results far from the truth\footnote{The KL divergence of $\infty$ is due to a floating-point underflow causing a $\log(0)$ term to appear in the calculation.} (determined by the large KL divergences), and all three take roughly 10 minutes to compute. Initializing with $\bb{\lambda^2}$ yields the relative best posterior estimate because the initial values were nearly equal to the true parameter values. Despite this, it still estimates many of the posterior parameters quite poorly. 

\subsection*{Hybrid CAVI with Numerical Optimization of ELBO}
To test the efficacy of Hybrid CAVI, we will use the mean components of the same three $\bb{\lambda}$ initial values as before, and see how Hybrid CAVI performs relative to the existing methods. In addition to comparing the results in Table \ref{fig:results3} to the results in Tables \ref{fig:results1} and \ref{fig:results2}, we can compare Hybrid CAVI's performance to the results from running a complete MCMC procedure. Note that both MCMC and Hybrid CAVI are using the Metropolis-Hastings MCMC algorithm (see \cite{bayesChoice} for details). 
\begin{table}[!htb]
\begin{center}
    \begin{tabular}{|c|c|c|c|c|}
    \hline
    Algorithm & Initialization & Time (s) & Estimated Posterior & KL Div. \\
    \hline
    MCMC & $\bb{\mu^1} = \mqty[10\\10]$ & 654.0 & $N_2\qty(\mqty[27.230 \\ 12.993], \mqty[0.418 & 0.0121 \\ 0.0121 & 0.0397])$ $\vphantom{\mqty[0\\0\\0]}$ & 71.128 \\
    & $\bb{\mu^2} = \mqty[25\\10]$ & 444.1 & $N_2\qty(\mqty[27.231 \\ 12.995], \mqty[0.352 & 0.00793 \\ 0.00793 & 0.0404])$ $\vphantom{\mqty[0\\0\\0]}$ & 69.979\\
    & $\bb{\mu^3} = \mqty[10\\20]$ & 541.7 & $N_2\qty(\mqty[27.222 \\ 12.983], \mqty[0.409 & 0.0135 \\ 0.0135 & 0.0427])$ $\vphantom{\mqty[0\\0\\0]}$ & 67.845\\
    \hline
    Hybrid & $\bb{\mu^1} = \mqty[10\\10]$ & 61.1 & $N_2\qty(\mqty[27.409 \\ 12.934], \mqty[0.288 & 0 \\ 0 & 0.0263])$ $\vphantom{\mqty[0\\0\\0]}$ & 95.750\\
    CAVI & $\bb{\mu^2} = \mqty[25\\10]$ & 56.2 & $N_2\qty(\mqty[27.153 \\ 13.027], \mqty[0.235 & 0 \\ 0 & 0.0263])$ $\vphantom{\mqty[0\\0\\0]}$ & 95.139\\
    & $\bb{\mu^3} = \mqty[10\\20]$ & 261.7 & $N_2\qty(\mqty[27.208 \\  13.031], \mqty[0.446 & 0 \\ 0 &  0.0263])$ $\vphantom{\mqty[0\\0\\0]}$ & 90.404\\
    \hline
\end{tabular}
\caption{Results for MCMC and Hybrid CAVI. Note that MCMC was run using 20,000 total steps with a 15,000 step burn-in. Hybrid CAVI used 1,000 total steps with a 900 step burn-in.}
\label{fig:results3}
\end{center}
\end{table}

The results presented in Table \ref{fig:results3} illustrate how it is much better to do Hybrid CAVI over the ordinary CAVI algorithm when limited to numerical optimization of the ELBO. The threshold for how many iterations of MCMC are necessary for Hybrid CAVI will be application-dependent, but there is clear merit in considering performing this MCMC burn-in. Not only are the parameter estimates closer to the truth, but we see a decrease in computation time by nearly a factor of 10 for Hybrid CAVI compared to both CAVI and MCMC, while yielding only a mild increase in KL divergence compared to a full MCMC procedure.

\subsection{Application to Non-Conditionally Conjugate Exponential Families}
\label{subsec:nonexp}
The above simulation demonstrates the advantages of Hybrid CAVI over MCMC and CAVI when closed-form updates are unavailable. Contrarily, it is obviously not useful to use Hybrid CAVI over CAVI when an analytic update exists. Thus, we demonstrate Hybrid CAVI on a multivariate $t$-distribution, which does not lie in a conditionally conjugate exponential family, and so no closed-form updates exist.

In our new setup, we generate 100 data points $\vb*{x}_1, \ldots, \vb*{x}_{100} \sim F_{X, Y}$, where $F_{X, Y}$ is a distribution function generated using a Gaussian copula with covariance matrix
\begin{equation*}
    \mqty[1 & 0.5 \\ 0.5 & 1]
\end{equation*}
and has marginals $F_X \sim t_8$ and $F_Y \sim t_{50}$. 
We then wish to estimate the degrees of freedom parameters $\nu_1$ and $\nu_2$ for the marginal distributions. To do so, following the advice of \cite{gelman2006}, we have an inverse uniform prior on $\nu_i$ given by $1/\nu_i \sim \operatorname{Uniform}(0, 1/2)$. For the CAVI algorithm, we then choose our variational factors $q_i(\nu_i; \vb*{\lambda}_i)$ to be shifted Gamma densities:
\begin{equation*}
    \nu_i - 2 \given \vb*{\lambda}_i \sim \operatorname{Gamma}(\lambda_{i, 1}, \lambda_{i, 2})
\end{equation*}
Because there is no closed form expression for the posterior, we take the posterior estimated by MCMC as the ground truth, and compare CAVI and Hybrid CAVI to the MCMC posterior. The KL divergence is then approximated using a discrete summation rather than the usual integral formulation (due to MCMC returning discrete sample points). Given marginal sample means $m_1$, $m_2$ and marginal sample variances $s_1^2$, $s_2^2$ of an MCMC chain, the method of moments estimators for $\vb*{\lambda}_i$ are given by $\lambda_{i, 1} = (m_i-2)^2/s_i^2$ and $\lambda_{i, 2} = s_i^2/(m_i-2)$. 

The initializations for MCMC and Hybrid CAVI are chosen to be the median vector on the prior ($\vb*{\nu^1} = (4, 4)$) and another arbitrary large starting point ($\vb*{\nu^2} = (30, 30)$). For CAVI, we choose a completely uninformed starting value ($\vb*{\lambda^1}$), an initialization that is very close to the optimal ($\vb*{\lambda^2}$), and an initialization that swaps the true optimal parameter values ($\vb*{\lambda^3}$).

\begin{table}[!htb]
\begin{center}
    \begin{tabular}{|c|c|c|c|c|}
    \hline
    Algorithm & Initialization & Time (s) & Approx. KL Div. \\
    \hline
    MCMC & $\vb*{\nu^1} = \mqty[4 \\ 4]$ $ \vphantom{\mqty[0\\0\\0]}$ & 33806 & 0 \\
    \hline
    Hybrid CAVI & $\vb*{\nu^1} = \mqty[4 \\ 4]$ $ \vphantom{\mqty[0\\0\\0]}$ & 1671 & $2.471$ \\
    & $\vb*{\nu^2} = \mqty[30 \\ 30]$ $ \vphantom{\mqty[0\\0\\0]}$ & 1557 & $2.915$ \\
    \hline
      CAVI & $\mqty[\vb*{\lambda^1}_1\\\vb*{\lambda^1}_2] = \mqty[(10, 10) \\ (10, 10)]$ $ \vphantom{\mqty[0\\0\\0]}$& 743& $\infty$ \\
    &$\mqty[\vb*{\lambda^2}_1\\\vb*{\lambda^2}_2] = \mqty[(3.2, 0.6) \\ (1.7, 6.9)]$ $ \vphantom{\mqty[0\\0\\0]}$& 553 & 0.788\\
    &$\mqty[\vb*{\lambda^3}_1\\\vb*{\lambda^3}_2] = \mqty[(1.7, 6.9) \\ (3.2, 0.6)]$ $ \vphantom{\mqty[0\\0\\0]}$& 577 & 9.081\\
    \hline
\end{tabular}
\label{fig:results4}
\end{center}
\caption{Results for all three posterior estimation techniques. Note that MCMC was run using 100,000 total steps with a 10,000 step burn-in. Hybrid-CAVI used 500 total steps with a 400 step burn-in.}
\end{table}

We once again see the advantages of Hybrid CAVI: The method achieves a small KL divergence in a fairly short amount of time relative to CAVI and MCMC respectively. Indeed, Hybrid CAVI is again much more stable in its accuracy compared to CAVI (only falling short of CAVI when CAVI is initialized close to the optimal values) and takes a fraction of the time of a full MCMC procedure. It is thus evident that Hybrid CAVI reaps the benefits exhibited by both CAVI and MCMC while mitigating their respective flaws.

\bibliographystyle{authordate1}
\bibliography{References}

\end{document}